\documentclass[pss,fleqn,embeddedheads]{w-art}

\usepackage{times}
\usepackage{w-thm}

\usepackage[]{graphicx}
\begin{document}

\DOIsuffix{theDOIsuffix} 
\Volume{XX} \Issue{1} \Copyrightissue{01} \Month{01} \Year{2004}
\pagespan{1}{} 
\Receiveddate{\sf zzz} \Reviseddate{\sf zzz} \Accepteddate{\sf
zzz} \Dateposted{\sf zzz} 

\title[Manuscript preparation guidelines]{The Fisher's information
and quantum-classical field theory: classical statistics similarity}

\author[J. Syska]{J. Syska
\inst{1}}
\address[\inst{1}]{Department of Field Theory and Particle Physics,
Institute of Physics, University of Silesia, \\ Uniwersytecka 4,
40-007 Katowice, Poland}

\begin{abstract}
The classical statistics indication for the impossibility to
derive quantum mechanics from classical mechanics is proved. The
formalism of the statistical Fisher information is used. Next the
Fisher information as a tool of the construction of a
self-consistent field theory, which joins the quantum theory and
classical field theory, is proposed.
\end{abstract}

\maketitle




\renewcommand{\leftmark}
{J. Syska: The Fisher's information and quantum models - classical
field theory, classical statistics similarity}

\section{Introduction}

It is said that classical mechanics is the stochastic limit of the
quantum mechanics\footnote{Yet it would be better to say that
classical mechanics has the symplectic (manifold) structure and
not the statistical one.}. And vice versa, as according to von
Neumann \cite{Neumann} quantum theory is incompatible with the
free of dispersion ensembles existence hence it is well recognized
that predicted departure from classical behavior of a system
appears at the statistical level only \cite{Peres}. Now, quite
recently the language of the geometry of the space of the
distributions has been formulated under the notion of statistical
geometry and the clue to its description is connected with the
Fisher information ($FI$) \cite{Fisher} matrix by which the
distance between distributions can be defined. It (Fisher-Rao
metric which is a Riemannian one) is used in the definition of the
relative entropy of two infinitesimally different
distributions\footnote{That is the infinitesimal version of the
Kullback-Leilbler relative entropy \cite{Kullback Pawitan}, which
is the tool for making the comparisons between the models,
especially in the time series analysis.}, being also the Hessian
matrix of the Shannon entropy \cite{Shannon,
Bengtsson_Zyczkowski}. Finally it is related to the notion of the
statistical $FI$ which (in the opposition to the global Shannon
entropy) characterizes local properties of the probability
distribution \cite{Frieden}. In the statistically orientated
theory of the measurement and estimation with the $N$ dimensional
sample, the $FI$ characterizes the local properties of the
likelihood function $p( {\bf y}; \; \theta_{1},..., \theta_{N} )$
which formally is the joint probability (density) of the data
${\bf y} \equiv ({\bf y}_{1}, ..., {\bf y}_{N})$ but is treated as
the function of parameters $\theta_{n}$. The set $\Theta =
(\theta_{1},..., \theta_{N})$ of the parameters\footnote{In the
most general case of the estimation procedure the dimensions of
${\bf y}$ and $\Theta$ are usually different.} are the coordinates
in the distribution space in which
the distance between distributions is defined. \\
During the course of the paper it will be shown that the
statistical $FI$ is a right tool to address two mutually related
problems. The first one (Section~2) is connected with the
statistical proof of the impossibility to derive quantum mechanics
from classical mechanics and the second one (Sections~2 and 3)
meets the consistency problem of the self-field formalism which is
used in such different branches of the physical research as the
superconductivity \cite{superconductivity}, atomic or particle
physics and astrophysics \cite{bib B-K-1}.

\section{The proof}

Below, in order to prove that quantum mechanics is not to be
derived from classical mechanics, the methods of quantum
mechanics and classical statistics are compared.\\
The maximum likelihood (ML) method is connected with the analysis
of the first derivative in the parameters of the $ln$-likelihood
function. Its second derivative with the minus sign leads to the
(observed) $FI$ \cite{Kullback Pawitan}. So, the relation between
the ML principle and the minimal statistical information (and
therefore maximal entropy (ME)) principle is not at all obvious.
Let ${\bf y}$ be the vector of position and $\widehat{\Theta} =
(\widehat{\theta}_{1},..., \widehat{\theta}_{N})$ the set of ML
estimates of the vector parameter $\Theta$.
In the case of two infinitesimally close distributions $p({\bf
y})$ and $p({\bf y} + \Delta {\bf y})$, the relation between the
(expected) $FI$\footnote{It could be proved that under the
regularity conditions the expected $FI$ is the variance of the
gradient $\partial ln p(\Theta)/\partial \Theta$ \cite{Kullback
Pawitan}. }, defined \cite{Kullback Pawitan} as
\begin{eqnarray}  \label{Fisher_information_1}
I \equiv - \sum_{n=1}^{N} \int d {\bf y} \frac{\partial^{2} ln
p}{\partial \theta_{n}^{2}} \; p({\bf y}) = \sum_{n=1}^{N} \int d
{\bf y} ( \frac{\partial ln p}{\partial \theta_{n}} )^{2} p({\bf
y}) \;
\end{eqnarray}
and the Kullback-Leibler entropy $G$ \cite{Kullback Pawitan} for
two infinitesimally different distributions, is as follows
\cite{Frieden}:
\begin{eqnarray}  \label{information-index}
I \sim  \, - \, G \left[ \, p({\bf y}), p({\bf y} + \Delta {\bf
y}) \right] \; .
\end{eqnarray}
Hence it is the notion of the (relative) entropy which is the
basic one and many of the properties of the entropy might be
rewritten into the language of information. When entropy is the
measure of disorder, the information is the measure of order.

Now, the quantum mechanical (q.m.) analog of the ML estimators are
the operators, i.e. observables, and the wave function which is
the basic quantity in the Schr{\"o}dinger quantum (wave) mechanics
is the carrier of the full information on the eigenvalues of these
operators. Parameters of (the distribution of) the random variable
are the analogs of the eigenvalues. Hence it might be noticed that
the quantum analog of the statistical random variable distribution
ought to be the quantum mechanical wave function.\\
The concussion follows that quantum mechanics is the statistical
methods. In statistics the sample is connected with the random
choice of some collection of {\it N} states from the whole
population of state collections (i.e. from the sample space). The
"most probable" values (estimates) for the population parameters
are those which maximize the joint probability of the data. This
procedure of choosing by maximization a particular distribution of
states is described under the name of the maximum likelihood
principle (MLp). Yet the ML method is not accomplished by the
maximization procedure of the likelihood as usually a lot of
distributions in the distribution space remain. Hence the finale
procedure is connected with further investigation of the shape of
the distribution \cite{Kullback Pawitan}, which is crucial, as
bigger the Fisher information (about the values of the parameters)
is the narrower the distribution is also. The question arises what
value should be chosen by the Fisher information during the
inquiry of the possible shape of the distribution but to answer it
a new criterion is needed. Proper physical models are to arise as
the consequences of a new information principle (IP). It will be
stated later on but already the feeling should be that the ME
principle (MEp) intervenes somehow. It should be stressed that
just mentioned IP (which should have the physical background)
might choose the estimators which are inefficient\footnote{In
connection to the Cramer-Rao inequality $e^{2} \, I \geq 1$ (where
$e^{2}$ is the mean-square error of the estimate from the true
value of the parameter), the {\it maximal} inverse of the possible
value of $e^{2}$ is called the {\it channel capacity} being equal
to the Fisher information.}. The flow diagram below summarizes the
discussed analog, on which inspection the question arises. Is e.g.
the Schr{\"o}dinger q.m. the consequence of the statistical IP, or
is it doing the choice of the distribution from the start by
yourself? {\small
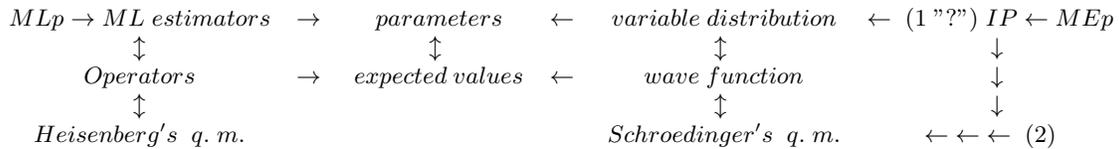
\begin{eqnarray*}
\begin{array}{cccccc}
  \!\!\!\!\!\!\!\!\!\!\!\!\!\! MLp \rightarrow ML \; estimators & \rightarrow &  \; parameters & \leftarrow &  \; variable \; distribution & \leftarrow \; (1 \; "?") \; IP \leftarrow MEp \\
  \!\!\!\!\!\!\!\!\!\!\!\!\!\! \updownarrow &  & \updownarrow &  & \updownarrow & \;\; \downarrow\\
  \!\!\!\!\!\!\!\!\!\!\!\!\!\! Operators & \rightarrow  &  \; expected \; values & \leftarrow &  \; wave \; function & \;\; \downarrow\\
  \!\!\!\!\!\!\!\!\!\!\!\!\!\! \updownarrow &  &  &  & \updownarrow & \;\; \downarrow\\
  \!\!\!\!\!\!\!\!\!\!\!\!\!\! Heisenberg's \;\; q.\; m. &  &  &  & \; Schroedinger's \;\; q.\; m.  & \leftarrow\; \leftarrow \; \leftarrow \; (2) \\
\end{array}
\end{eqnarray*}}
Furthermore, analyzing the form of the statistical $FI$ of the
system, it could be noticed \cite{Frieden} that it may be written
in real amplitudes $q_{n}$ as follows:
\begin{equation}  \label{Fisher_information}
I = 4 \sum_{n=1}^{N} \int d {\bf x}_{n} \left( \frac{\partial
q_{n}}{\partial {\bf x}_{n}} \right)^{2} \; , \;\;\;\; {\rm where}
\;\;\;  q_{n}^{2} ({\bf x}_{n}) \equiv p_{x_{n}} ({\bf x}_{n})\; .
\end{equation}
Here $p_{x_{n}} ({\bf x}_{n})$ is the probability
distribution\footnote{With the substantial or probabilistic
origin. From Section~3 it follows that as the substantial origin
is appropriate e.g. for the Maxwell electro-magnetic field, it is
then appropriate for the Dirac wave function also as both of them
have the same statistical origin.} which has the property of the
{\it shift invariance} i.e., $p_{x_{n}} ({\bf x}_{n})$ $=
p_{x_{n}} ({\bf x}_{n}|\theta_{n}) = p_{n} ({\bf
y}_{n}|\theta_{n})$ with ${\bf x}_{n} \equiv {\bf y}_{n} -
\theta_{n}$, where $(\theta_{n})$ is the set of physical
quantities (parameters with unknown values) of a physical nature
(e.g. positions); ${\bf y}_{n}$ are $N$ data values and ${\bf
x}_{n}$ are added displacements (fluctuations). The transition
from Eq.(\ref{Fisher_information_1}) to
Eq.(\ref{Fisher_information}) is performed under the
assumption\footnote{The chain rule $\partial/\partial \theta_{n} =
\partial/\partial ({\bf y_{n}} - \theta_{n}) \;\; \partial ({\bf
y_{n}} - \theta_{n})/\partial \theta_{n} = - \; \partial/\partial
({\bf y_{n}} - \theta_{n}) = - \; \partial/\partial {\bf x_{n}}$
has also been used.} that the data are collected independently
which allows to express the joint probability $p({\bf y})$ via the
{\it factorization property} as $p({\bf y}) \equiv p({\bf
y}|\Theta) = \prod_{n=1}^{N} p_{n}({\bf y}_{n}|\theta_{n})$, where
$\theta_{m}$ has no influence on ${\bf y}_{n}$ for $m \neq n$. Now
from amplitudes $q_{n}$ the wave function $\psi_{n}$ could be
constructed as follows:
\begin{equation}  \label{psi}
\psi_{n} = \frac{1}{\sqrt{N}} (q_{2 n - 1} + i \, q_{2 n}) \; ,
\;\;\;\; {\rm where} \;\;\; n = 1, 2, ..., N/2 ,
\end{equation}
with the number of real degrees of freedom being twice the complex
ones. After using the total probability law for all data, the
probability distribution for the system (e.g. a particle) could be
rewritten as $p({\bf x})$ $= \sum_{n=1}^{N} p_{x_{n}} ({\bf
x}_{n}|\theta_{n}) P(\theta_{n}) = \frac{1}{N} \sum_{n=1}^{N}
q_{n}^{2}$, where we have chosen $P(\theta_{n}) = \frac{1}{N}$
according to our lack of knowledge on which one of $\theta_{n}$
actually occurs in the $n$-th experiment\footnote{$P(\theta_{n}) =
\frac{1}{N}$ has nothing to do with the Bayesian distribution of
one particular parameter $\theta_{n}$,
but with the random choice of any one of it from the set
$(\theta_{n})$ during the course of the evolution of the system.
In this respect the graphical interpretation of the Feynman path
integrals might be useful.}. All of these leads to
\begin{eqnarray}  \label{total_probability-psi} p({\bf x})
= \sum_{n=1}^{N/2} \psi_{n}^{*} \psi_{n} \;
\end{eqnarray}
which establishes the right relation between the probability and
the wave function. Let us notice that the shift invariance
condition together with the factorization property are very
important. Under them the information $I$ does not depend on the
parameter set $(\theta_{n})$ \cite{Kullback Pawitan}, and the wave
functions (\ref{psi}) do not depend on these parameters (e.g.
positions) also. The distribution (probability law
(\ref{total_probability-psi})) is then of the form $p({\bf
x})=|\psi({\bf x})|^2$ rather than $|\psi({\bf x}|\Theta)|^2$
\cite{Frieden}. Finally  the $FI$ (\ref{Fisher_information}) could
be explicitly rewritten in the shape of the kinetic action term
\begin{equation}  \label{kinetic_action}
I = 4 \, N \sum_{n=1}^{N/2} \int d{\bf x} \, \frac{\partial
\psi_{n}^{*}({\bf x})}{\partial {\bf x}} \; \frac{\partial
\psi_{n}({\bf x})}{\partial {\bf x}} \; ,
\end{equation}
where index $n$ has been dropped from the integral as the range of
all ${\bf x}_{n}$ is the same. In this way it was proven
\cite{Frieden} (at least at this stage) that quantum mechanics is
indeed a statistical method, yet notice that the interpretation of
the wave function as the classical (i.e. real) probability
distribution is misleading, as in Eq.(\ref{psi}) we have finished
with the complex wave function $\psi_{n}$.

Until now we have not established the value of $N$, the sample
size. In general, if $N \rightarrow \infty$ the ML estimators have
elegant properties, they are unbiased end efficient. In classical
mechanics to specify precisely the parameter (e.g. the position of
the point-like particle) means that the infinite number $N$ of
experiments should be carried out. Hence in classical mechanics
$N$ in Eq.(\ref{Fisher_information}) goes to infinity.  It has
been also proved that the quantum mechanical models are obtained
with the notion of the $FI$ for the precise, finite values of $N$
(see \cite{Frieden}). E.g. for the Klein-Gordon\footnote{The
formalism of the $FI$ might be easily generalized leading to the
relativistically covariant equations. If e.g. ${\bf x_{n}}$ is a
vector $(x_{n}^{\nu})$ then we introduce in
Eq.(\ref{Fisher_information}) the notation $(\frac{\partial
q}{\partial {\bf x_{n}}})^{2} \equiv \sum_{\nu} \frac{\partial
q}{\partial x_{n \, \nu}} \frac{\partial q}{\partial x_{n}^{\;
\nu}}$ ; and the same notation for $\theta_{n} = (\theta_{n}^{\;
\nu})$.} and Schr{\"o}dinger equation (as its limit) $N = 2$, for
Dirac equation $N=8$, for Maxwell equations $N=4$. Frieden derived
also the classical mechanics from the quantum model but as the
limit case $\hbar \rightarrow 0$ only and it could be noticed that
the value of $N$ is irrelevant in his calculations. Now we see
that models belong to the different cases of $N$. For the
particular quantum or classical field model $N$ is finished (as
only eigenvalue is needed) but for classical mechanics $N$ should
be infinite (to have the information on the position of the
classical particle at every moment of time).\\
Suppose that we have a system which is described by a nonsingular
distribution. Then for $N \rightarrow \infty$ the $FI$
(\ref{Fisher_information_1}) diverges to infinity. Yet the same
happens for any singular distribution like the Dirac delta
distribution also. To see it let us consider a point-like free
particle at rest at the position $\theta$ and take a
$\delta$-Dirac sequence of functions, e.g. the sequence of the
Gauss functions $\delta_{k}({\bf y}_{n}) = \frac{k}{\sqrt{\pi}} \,
exp(- k^{2} ({\bf y}_{n} - \theta)^{2})$. Then, because for the
particular index $k$ the $FI$ is equal to
$\frac{N}{\sigma_{k}^{2}}$ \cite{Kullback Pawitan}, where
$\sigma_{k}^{2} = \frac{1}{2 k^{2}}$ describes the variance of the
position of the particle for the $k$-th element in the sequence,
we see that the $FI$ diverges to infinity for $N \rightarrow
\infty$ (and even more for $k \rightarrow \infty$). To sum up, for
$N \rightarrow \infty$ the $FI$ does not exist whatever the
distribution would be.\\
Hence there are two classes of theories pertaining to the
dimension $N$ of the sample, i.e. $N$ for the quantum mechanics
(and classical field theory also) is {\it finite} whereas for the
classical mechanics it is {\it infinite} which means that the
classical mechanics has not the statistical origin. This has
finished the proof that there is the inherent difference between
quantum and classical mechanics. To my best knowledge it has not
yet been given in this simple statistical form. The proof does not
encompass the impossibility of the derivation of quantum mechanics
(or other quantum theory) from a
classical {\it field} theory (or self-consistent field theory). \\
Let us notice that $FI$ looks like the action for the kinetic
energy term (Eq.(\ref{kinetic_action})). Using this quantity (and
new postulates on the physical information during the process of
the measurement, Section~3), Frieden \cite{Frieden} derived some
of the quantum mechanical models via the way (2) from the above
flow diagram.

\section{Fisher information and the self-field theory}

The result which follows from previous Section is that all
physical models fall into two categories. They are of the
classical mechanics origin or of the statistical one. So, the
division does not lie between what is micro or macro but what is
of the statistical or classical mechanics origin, still better,
what is of the field (wave) theory or strictly point-like origin.
The consequences are as follows. Mixing classical mechanics with
field theory models leads to the inconsistency as the one for the
Lorentz-Abraham-Dirac equation which leads to the self
acceleration of point-like charged particle which interacts with
its own electromagnetic field \cite{Rohrlich}. From the other side
combining quantum (wave) mechanics with classical electrodynamics
is more promising.\\
The question arises, are both wave (quantum) mechanics and
classical field theories of the statistical origin?
According to Section~2 there is a reason to acknowledge any
quantum model, which has the resemblance of its kinetic action to
the $FI$, as the statistical one (see also
\cite{Bengtsson_Zyczkowski}). It has been also shown
\cite{Frieden} that the main classical model, that is Maxwell
electrodynamics, has the same statistical structure. At this point
we need the construction of the statistical predecessors for both
the {\it kinetic action} and the {\it structural one}. From the
statistical perspective the first one is {\it the carrier of
information about the system in the measurement} but the second
one is {\it the carrier of information about the structure of the
system which reveals itself somehow in the measurements scenario}
taking into account additional constraints \cite{Frieden}. We need
to bind both types of information by the {\it new principle}. As
it has been said in Section~2, the predecessor for the kinetic
part is the Fisher information $I$. The construction of the
structural statistical term, called $Q$, follows the particular
characteristics of the theory which take into account the physical
parameters of the model. According to
Eq.(\ref{Fisher_information}), $I$ is the function of the
amplitudes $q(x)$, so $Q$ has to be also. Yet $Q$ has to depend on
the physical constants of the particular scenario also, e.g. on
$\hbar$ or $c$. Now, because there exists the entropy for the
kinetic term, namely the Kullback-Leibler relative entropy $G$
with the implication $G \rightarrow I$
(Eq.(\ref{information-index})) then there should exist the entropy
term for $Q$ also, let us call it $S_{Q}$.\\
A system which is without a structure dissolves itself hence its
equation of motion requires a structural term and "during putting
this structure upon" the entropy of the system has to be minimized
and information maximized. {\it Yet when} the constraints had been
established then from all distributions the one which maximizes
the entropy and minimizes information should be chosen via a {\it
new variational principle} $G + S_{Q} \rightarrow max$  or $\, I +
Q \rightarrow min \, $ which we call the {\it scalar principle}.
It might be written in the following form
\begin{eqnarray}  \label{principle_1}
\delta( I + Q ) = 0 \; , \;\;\; {\rm principle \; I \;\;\;
(scalar)}
\end{eqnarray}
and interpreted as a conservation law of the physical information
$K \equiv I + Q$ of the system. Both $I$ and $Q$ are (final)
information which exist in the system but only $I$ reveals to the
observer in the process of the measurement. Although $Q$
influences $I$, it is lost to the observer (in the measurement)
and is carried inside the system only. The first principle
(\ref{principle_1}) does not exhaust all possibilities. The
intriguing thing is that a lot of calculations might be done for
the most pessimistic scenario under which the total entropy
partitions itself equally (or with a factor 1/2) into the Fisher
and structural parts, having in total the value zero. Hence in
practice it occurred \cite{Frieden} that the law
(\ref{principle_1}) has to be completed by the following one:
\begin{eqnarray}  \label{principle_2}
I + \kappa \;Q  = 0 \; , \;\;\; {\rm where} \;\;\; \kappa = 1
\;\;\;{\rm or} \;\;\; 1/2 \; , \;\;\; {\rm principle \; II \;\;\;
(internal) \; ,}
\end{eqnarray}
which we call the {\it internal principle}. The minus sign of $Q
\sim - I$ is not so strange as it seems. For example for a pure
classical state the Shannon entropy goes to minus infinity and
this means that an infinite amount of information should be taken
to specify such a state exactly \cite{Bengtsson_Zyczkowski}. \\
In \cite{Frieden} the other approach to the structure information
was presented. Frieden introduced the so called bound information
$J$ which has the interpretation of being confined in the system
before the measurement. Although Frieden axioms are operationally
similar to Eq.(\ref{principle_1}) and (\ref{principle_2}) if only
$J = - Q$, yet the difference in the interpretations is obvious.
As the system in Frieden interpretation exhibits during the
measurement the transfer of information $I \rightarrow J$, having
at any moment of time one of these two types of information only,
in our scenario the system is characterized by $I$ and $Q$
simultaneously at any moment of time. \\
At first look (it has lasted over 70 years) it seems that the
Kline-Gordon and Dirac equations are more similar to each other
than the Dirac and Maxwell ones. But, using Eq.(\ref{principle_1})
and Eq.(\ref{principle_2}) the Klein-Gordon equation is obtained
whereas Eq.(\ref{principle_2}) alone gives the Dirac equation or
Maxwell equations for $\kappa = 1$ or $\kappa = 1/2$, respectively
\cite{Frieden}. Hence the Dirac and Maxwell cases are more similar
in their axiomatic origin. Yet the source of their difference in
the $\kappa$ value is also important. The comparison of the cases
of Dirac and Maxwell equations suggests that the ratio $Q$ to $I$
is in the Maxwell case twice as big as in the Dirac case. In this
context one more puzzle is solved. In 1990 Sallhofer
\cite{Sallhofer Sakurai_2} completed the model of the (hydrogen)
atom, based on the isomorphism between Maxwell and Dirac
formalisms. He, in the Minkowski space, worked out the formal
mathematical strong similarity (I do not call it identity) of
electrodynamics and wave mechanics by means of which he proved
that the hydrogen atom might be seen as the pair of mutually
refracting electromagnetic waves. Previously this similarity was
pointed out by Sakurai \cite{Sallhofer Sakurai_2}. Starting from
the Maxwell equations Sallhofer obtained as if the Dirac equation
for the hydrogen atom but with twice as much components for the
"electronic" field than there are in the original Dirac equation.
The physical structural identification of some of these components
gives four degrees of freedom, as for the Dirac field
\cite{Sallhofer Sakurai_2}, which means that the Maxwell equations
are of more fundamental nature than the Dirac one.

After the choice of the axiom I or II (which one to choose should
be verified in the experiment), the calculations of $Q$ which
follow are sometimes tedious. The simplest case exists for the
scalar particle with $N = 2$ (see \cite{Frieden}). So, we have the
single complex wave function $\psi({\bf x})$ in the position ${\bf
x}$ space and its Fourier transform $\phi({\mathbf \mu})$ in the
momentum ${\mathbf \mu}$ space.  After choosing the {\it internal}
principle (\ref{principle_2}) with $\kappa = 1$, $\,I\left[
\psi_{0}({\bf x}) \right] + Q\left[ \phi_{0}({\mathbf \mu})
\right] = 0 \, $, which means that information is equally
distributed among the kinetic and structural parts, the Fisher
information $I$ and the structural information $Q$ are equal to $
I\left[ \psi \right]~=~8 \int d{\bf x} \frac{\partial
\psi^{*}({\bf x})}{\partial {\bf x}} \; \frac{\partial \psi({\bf
x})}{\partial {\bf x}} \;$ and $\, Q\left[ \phi \right] = - \,
\frac{8}{\hbar^2}\int d {\mathbf \mu} \, \mu^2 \;
\phi^{*}({\mathbf \mu}) \; \phi({\mathbf \mu})$, respectively. The
wave functions $\psi$ and $\phi$ satisfying this internal
principle are $\psi_{0}$ and $\phi_{0}$, respectively. Yet to
obtain the Klein-Gordon equation the {\it scalar} principle
(\ref{principle_1}) should be used also. But, whether for the
scalar, spinor or vector field\footnote{To obtain the Dirac or
Maxwell equations the internal principle II is enough.} the
proposed procedure leads to the proper information $I$ and $Q$,
giving in the result kinetic and structural actions and in the
result equations of motion. This fact means that there is the
statistical quantity (namely information) which precedes action
and that there are the information principles (I or II) which
stand before the variational principle of the total action.

\section{Conclusions}

In the paper the classical statistics proof of the impossibility
to derive quantum mechanics from classical mechanics has been
presented. This statement has appeared as the conclusion from the
fact that the Fisher information for different cases of the
dimension $N$ of the finite sample gives different field theories,
hence none of them is equivalent to classical mechanics for which
$N$ is infinite. Physically it might be understood as the result
of the fact that for the particular quantum model with $N$
established, (in order to describe the state) the eigenvalue is
needed only, but to have information on the position of the
classical mechanics particle at every moment of time, infinite $N$
is needed. To obtain any field theory two new principles were
proposed, the scalar one connected with the minimization of total
information in the system and the internal one on zeroing this
total information. It was pointed out that the notion of
information stands before the usual physical action. Much of the
work has been done previously by Frieden and Soffer \cite{Frieden}
and it rightly might be called the Frieden approach to equations
of motion, yet the method should be reinterpreted, particularly in
understanding the structural information and developed in finding
the information predecessors for sources and physics
beyond the value of $N$. \\
Finally, because for the construction of different field theory
models (classical and quantum) the same formalism of statistical
Fisher information has been used hence it is the tool to the
construction of a self-consistent field theory also \cite{bib
B-K-1}, the one which joins the quantum theory and classical field
theory in one logically consistent mathematical apparatus.

\section*{Acknowledgments}
This work has been supported by L.J.Ch..\\
This paper has been also supported by the Polish Ministry of
Scientific Research and Information Technology under the
(solicited) grant No PBZ-MIN-008/P03/2003 and by the Department of
Field Theory and Particle Physics, Institute of Physics,
University of Silesia.

\end{document}